# Specific Heat of $CeRhIn_5$: Pressure-Driven Evolution of the Ground State from Antiferromagnetism to Superconductivity

R. A. Fisher,[1] F. Bouquet,[1] N. E. Phillips,[1] M. F. Hundley,[2] P. G. Pagliuso,[2] J. L. Sarrao,[2] Z. Fisk,[3] and J. D. Thompson[2]

[1]*Lawrence Berkeley National Laboratory and*
*Department of Chemistry, University of California, Berkeley, CA 94720*
[2]*Los Alamos National Laboratory, Los Alamos, NM 87545*
[3]*NHMFL, Florida State University, Tallahassee, FL 32306*

Measurements of the specific heat of antiferromagnetic $CeRhIn_5$, to 21 kbar, and for 21 kbar to 70 kOe, show a discontinuous change from an antiferromagnetic ground state below 15 kbar to a superconducting ground state above, and suggest that it is accompanied by a weak thermodynamic first-order transition. Bulk superconductivity appears, apparently with d-wave electron pairing, at the critical pressure, 15 kbar; with further increase in pressure a residual temperature-proportional term in the specific heat disappears.

PACS numbers: 74.25.Bt, 74.70.Tx, 75.50.Ee, 75.40.Cx

The occurrence of superconducting (SC) heavy-fermion (HF) compounds provides a unique opportunity for investigating the relation between magnetism and superconductivity, particularly the possibility of magnetically mediated pairing of the electrons. In magnetic HF compounds there is a competition between magnetic order, driven by the RKKY interaction, and the spin-singlet ground state, favored by the Kondo interaction [1]. These interactions are both governed by the localized-moment conduction-electron exchange $|J|$, but the dependence on $|J|$ is different, quadratic for the RKKY, and exponential for the Kondo interaction. Since $\partial|J|/\partial P > 0$, the application of pressure ($P$) can reduce the magnitude of the ordered magnetic moments and lower the temperature of the ordering. For appropriate values of the relevant parameters the Néel temperature ($T_N$) of an antiferromagnetic (AF) HF compound, and the Curie temperature ($T_C$) of a ferromagnetic (FM) HF compound can be driven to zero at a critical pressure ($P_c$). There has been considerable speculation that superconductivity might appear in that limit, with the electron pairing mediated by the strong magnetic fluctuations associated with the magnetic-nonmagnetic boundary, but the number of likely examples is small, presumably because the conditions that must be satisfied for superconductivity to be realized are so restrictive. Those conditions and the relevant concepts have been summarized in the context of the observed superconductivity in AF $CePd_2Si_2$ and $CeIn_3$, and a general phase diagram proposed [2]. One of the conditions emphasized is that the magnetic transition must be "continuous" at $P_c$. A phase diagram with similar features has been reported [3] for FM $UGe_2$, supporting the general picture outlined in Ref. [2]. For all three of these materials the critical temperature for magnetic ordering approaches zero at $P_c$, and superconductivity appears in a narrow window of $P$ in the vicinity of $P_c$, with a strongly $P$-dependent critical temperature ($T_c$) that is a maximum in the vicinity of $P_c$. However, in a recent Letter, Hegger *et al.* [4] have reported a transition from magnetism to superconductivity and a phase diagram of a very different form for AF $CeRhIn_5$. Their phase diagram, based on measurements of the resistivity ($r$), is represented in Fig. 1. $T_N$, ~ 4 K, is only weakly dependent on $P$ to 14.5 kbar; at the next higher $P$, 16.3 kbar, the signature of AF order has disappeared, and superconductivity has appeared with an essentially $P$-independent $T_c$ of ~ 2 K. This abrupt change suggested a "first-order-like transition" at $P_c$ ~ 15 kbar. (Data points below and above $P_c$ are distinguished by solid and open symbols, respectively, in all figures.)

The phase diagram for $CeRhIn_5$ and its implications for the relation between magnetism and superconductivity suggest further study. Furthermore, the features of interest occur in regions of $P$ and $T$ that are relatively accessible to measurements of the specific heat ($C$), which gives relevant information that is not obtained from the more usual measurements of resistivity and magnetization. In this Letter we report measurements of $C$ that show that the superconductivity indicated by the measurements of $r$ is a bulk phenomenon. They characterize the superconductivity, and determine the nature of the low-energy excitations, thereby identifying the ground states throughout the range of $P$. The ground states, AF below $P_c$ and HF/SC above, both evolve continuously with increasing $P$, but at $P_c$ there is a discontinuous change: Long-range AF order disappears and superconductivity appears. The data suggest that the change is accompanied by a weak thermodynamic first-order transition.

The sample, consisting of small, randomly oriented crystals, was contained in a clamped pressure cell with AgCl as the pressure-transmitting medium, and Sn and Pb manometers, one at each end of the sample. The pressure differences between the two ends of the sample were at most 0.2 kbar.

The zero-field measurements of $C$ are shown in Fig. 2 for representative values of $P$. The lattice heat capacity ($C_{\mathrm{lat}}$), taken to be the same as that of $LaRhIn_5$ [4], shown in Fig. 2, was subtracted from $C$ to obtain the "electron" contribution ($C_e$), shown on an expanded scale in Fig. 3. For all $P$ the data permit plausible extrapolations to $T = 0$, and the entropy ($S_e$) calculated at 12 K has the same value for all $P$ to within ± 2 %.

Characteristic temperatures derived from $C_e$ and $r$ are compared in Fig. 1. With increasing $P$ the specific-heat anomaly that is associated with AF ordering at ambient pressure becomes broadened and reduced in amplitude. The temperature of the maximum in $C_e/T$ ($T_{\max}$) tracks the $T_N$ deduced from $r$ (including the small increase at low $P$) for $P \leq 10$ kbar, but then shifts to lower $T$. (At 12 kbar $T_N$ determined by NQR measurements [5] is close to $T_{\max}$.) For $P \geq P_c$, $T_{\max}$ is close to the unidentified feature [4] in $r$ at $T_?$. (Evidently $r$ detects two features in the 9 - 15 kbar region that are not resolved in $C_e$.) A second anomaly, associated with the transition to the SC state, first appears as a small irregularity at 16 kbar (in data that are too close to those at 16.5 kbar to be included in Fig. 2). It grows to a "shoulder" on the larger anomaly at 16.5 kbar and reaches its maximum amplitude at 19 kbar. Values of $T_c$, taken as the midpoints of entropy-conserving constructions on $C_e/T$, are in good agreement with the values determined from $r$, which correspond to the onset of superconductivity.

For $P = 21$ kbar and $H = 0$, 50, and 70 kOe, the values of $T_c(H)$ conform to a parabolic $T$ dependence of $H_{c2}(T)$ that extrapolates to $H_{c2}(0) = 159$ kOe. Apart from the extrapolation, there is an uncertainty in this value of ~ 20 kOe, but it is in satisfactory agreement with the value 152 kOe, obtained from a similar treatment [4] of resistivity data. For $T < T_c(H)$, $C_e(H) = g(H)T + B_2(H)T^2$, as shown in Fig. 4. Extrapolations of $C_e(H)/T$ to 0 K (see insets) give the same $S_e$ at $T_c(0)$, $S_e(T_c)$, to within ± 1 %, supporting the validity of both the extrapolations and the value of $S_e(T_c)$. As shown in Fig 5(b), $g(H)$ is proportional approximately to $H$, and extrapolation to $H_{c2}(0)$ gives $g$ = 382 mJ $K^{-2}$ $mol^{-1}$ for the normal-state value at 21 kbar. (For internal consistency the values of $g$ and some other parameters are given to more significant figures than warranted by the data.) The dependence of $C_e$ on $T$ and $H$ is similar to that of, e.g., $URu_2Si_2$ [6]. The superconducting-state $C_e$ ($C_{es}$) is $C_{es} = B_2(0)T^2$, which is characteristic of a certain group of heavy-fermion superconductors that also includes, e.g., $UPt_3$ [7], and is associated with an "unconventional" order parameter and line nodes in the energy gap [8]. The corresponding power-law $T$ dependence for nuclear-spin relaxation times [8] has been seen [5, 9] in NQR measurements.

At 21 kbar the normal-state $C_e$ ($C_{en}$) is defined to within narrow limits: For $T > T_c(H)$, $C_{en}$ is independent of $H$ and determined by the 70-kOe data to 1.7 K, as shown in Fig. 4. The interpolation between 1.7 K and 0 K, where $C_{en}/T$ = 382 mJ $K^{-2}$ $mol^{-1}$, must give the same $S_e(T_c)$ as that given by the low-$T$ data for $H$ = 0, 50, and 70 kOe. The curve in Fig. 4 is an almost-unique, plausible interpolation that satisfies these conditions. It has a shape similar to that of some other HF compounds, e.g., $URu_2Si_2$ [6] and $CeAl_3$ [10]. The discontinuity in $C_e$ at $T_c$ is relatively small: $\Delta C_e(T_c)/C_{en}(T_c) = [C_{es}(T_c) - C_{en}(T_c)]/C_{en}(T_c)$ is 1.43 for a BCS superconductor and ~ 1 to 1.5 for a number of HF superconductors, but only 0.36 for $CeRhIn_5$. However, the small value is a direct consequence of the $T$ dependences of $C_{es}$ and $C_{en}$, and requires no independent microscopic interpretation: If $C_{en} = gT$ with $g$ constant, and $C_{es} \propto T^2$, equality of entropies at $T_c$ requires that $\Delta C_e(T_c)/C_{en}(T_c) = \Delta C_e(T_c)/gT_c = 1$. For $CeRhIn_5$, as for many other HF superconductors, the density of quasiparticle states is energy-dependent, and $C_{en}$ must be represented by a $T$-dependent $g$, defined by $C_{en}(\mathrm{T}) \equiv g(T)T$. In that case, $\Delta C_e(T_c)/C_{en}(T_c) = 2S_{es}(T_c)/C_{en}(T_c) - 1$, where $S_{es}(T_c)$ is the superconducting-state entropy at $T_c$. $CeRhIn_5$ is evidently a somewhat extreme case in which $g$ is still strongly $T$ dependent at $T_c$, but it is not qualitatively different from, e.g., $URu_2Si_2$ for which the deviation of $g(T)$ from $g(T_c)$ is less precipitous and only 20% at 0 K, and $\Delta C_e(T_c)/C_{en}(T_c)$ ~ 0.9 [6]. At 21 kbar, the zero value of $g(0)$ shows that the Fermi surface (except for the line nodes) is fully gapped. For $T \leq T_c(H)$, $C_e(H)$ and $S_e(H)$ conform to expectations for superconducting material, and any additional contributions to $C_e$ must be negligible. By that criterion, the superconductivity at 21 kbar is complete as well as bulk.

Although in the vicinity of $T_{\max}$ the anomaly in $C_e$ evolves with increasing $P$ without discernable discontinuity, the $T$ dependence of $C_e$ at low $T$ is discontinuous at $P_c$, as is apparent in Figs. 2(a) and 3 where $C_e/T$ shows positive curvature for $P < P_c$, but zero curvature for $P > P_c$ as $T \rightarrow 0$. For all $P$, the lowest-order term in $C_e$ is $g(H)T$. For $P < P_c$, the second term is $B_{\mathrm{AFSW}}(H)T^3$ (Fig. 2(c)), which represents the spin-wave contribution expected for an antiferromagnet; for $P > P_c$, it is $B_2(H)T^2$ (Fig. 2(b)), which, as noted above, is characteristic of certain heavy-fermion superconductors. With increasing $P$ $B_{\mathrm{AFSW}}(0)$ increases monotonically through the AF region, corresponding to a progressive decrease in the spin-wave stiffness, i.e., in the product of the moment and the exchange interaction, before that term disappears abruptly in the vicinity of $P_c$. At $P_c$, $B_{\mathrm{AFSW}}(0)T^3$ is replaced by the superconductivity-related term, $B_2(0)T^2$, the coefficient of which then increases monotonically through the SC region. The discontinuous behavior of $g(0)$ is displayed in Fig. 5(a), which shows the experimental AF values with a plausible interpolation to the 21-kbar, normal-state value on the SC side, and the experimental SC values. The SC values extrapolate to the AF curve at ~ 15 kbar, the value of $P_c$ deduced from resistivity measurements [4]. In fields great enough to suppress superconductivity, $g$, which measures the density of low-energy excitations, would increase monotonically to the normal-state value at 21 kbar. In zero field, a transition to the superconducting state sets in at $P_c$, leaving a diminished, “residual” $g(0)$ that goes to zero at 21 kbar.

At $P_c$ there is no discontinuity in $g(0)$. On the SC side of the phase boundary $g(0)$ is the same in the superconducting and normal states and $\Delta C_e(T_c) = 0$. With increasing $P$ $g(0) \rightarrow 0$ and $\Delta C_e(T_c)$ increases, but with essentially no increase in $T_c$. The extended gapless regions on the Fermi surface of superconductors with $d_{x^2-y^2}$ pairing [12] provides a possible basis for understanding this behavior. Below a critical value of the pairing potential the gap vanishes and there is a finite density of low-energy quasiparticle states and $g(0) \neq 0$. With increases in the pairing potential the gap appears and increases in amplitude, and the quasiparticle density of states decreases. For sufficiently high gap amplitudes, the Fermi surface is "fully gapped" and the quasiparticle density of states approaches zero. The observed relation between $\Delta C_e(T_c)$ and $g(0)$ would correspond to an increase in the gap amplitude and pairing potential with increasing $P$.

Isotherms, $S_e(P)$ vs $P$, obtained by integration of $C_e(T)/T$ to obtain $S_e(T)$ and interpolation to fixed $T$'s, reveal interesting features near 12 and 15 kbar (see Fig. 6). The volume thermal expansion is proportional to $(\partial S_e/\partial P)_T$, but opposite in sign. Although the isotherms show that the thermal expansion is negative in most of the range of $P$ and $T$, they are consistent with the positive thermal expansion reported [11] at ambient pressure. The features at 12 and 15 kbar are emphasized in Fig. 6 by three straight-line segments that also represent thermodynamic transitions, respectively, second- and first-order. For any one isotherm the structure at 12 and 15 kbar represented by the straight lines is comparable to the deviations of the points from any smooth curve that might be drawn as an approximate fit to all the points, but its systematic variation from one isotherm to the next attests the reality of structure at least qualitatively similar to that represented by the lines. The feature near 12 kbar, the better defined of the two, is a relatively clear discontinuity in $(\partial S_e/\partial P)_T$, which corresponds to a discontinuous increase in the magnitude of the (negative) thermal expansion. It is a maximum near 3.5 K, and near zero at both 0.5 and 4.5 K. It marks a second-order transition that occurs at 12.0 and 11.2 kbar at 0.5 K and 4.5 K, respectively—the region of the phase diagram in which unusual features in the resistivity and susceptibility have been observed [4], and where $T_{\max}$ starts to deviate from its low-$P$ value. With the slope of the phase boundary and the Ehrenfest relation, it gives a maximum discontinuity in $C$ of ~50 mJ $K^{-1}$ $mol^{-1}$. The experimental data do not permit a meaningful quantitative comparison, but they are not inconsistent with that value. The feature at 15 kbar is less well defined, but the points above and below 15 kbar cannot be connected by smooth curves without a change in sign of the curvature. As represented by the straight lines, they suggest a finite $\Delta S_e$ at $P_c$ = 15 kbar, which is a maximum of ~ 0.15 J $K^{-1}$ $mol^{-1}$, 0.025Rln2, near 3.5 K and decreases to near zero at 0.5 and 4.5 K. Clearly, the points are not sufficiently closely spaced in $P$ to define precisely the interval in which this change in $S_e$ takes place, but there is certainly some small irregularity in $S_e$ of at least similar form in the vicinity of $P_c$. The straight lines represent a first-order transition from a low-$P$ phase, which must have the larger volume, to a high-$P$ phase that has a lower $S_e$. The sign of $\Delta S_e$ is consistent with the disappearance of the spin-wave contribution to $C_e$ at $P_c$. A quantitative comparison of magnitudes is limited by the uncertainty in $\Delta S_e$ and the extrapolation of $C_e$ to $P_c$, but, for $T \leq 3$ K, i.e., near and below the region of its maximum, $\Delta S_e$ is a factor of 2 - 5 smaller than the entropy of the spin-wave term at 13 kbar.

In conclusion, we have shown that with increasing $P$ the specific-heat anomaly associated with AF ordering in $CeRhIn_5$ at ambient $P$ evolves into a form more typical of a spin-singlet HF compound at 21 kbar. There is a discontinuous change in the nature of the ground state, from one of AF order to that of a SC HF compound at $P_c$ ~ 15 kbar. Our results confirm the major features of the phase diagram [4] that distinguish it from those of related compounds [2, 3], but show that, in spite of the abrupt disappearance of AF order at $P_c$, the magnetic transition there has only a small thermodynamic first-order component.

We are grateful to A. V. Balatsky and V. Z Kresin for helpful discussions. The work at LBNL was supported by the Director, Office of Basic Energy Sciences, Materials

Sciences Division of the U. S. DOE under Contract No. DE-AC03-76SF00098. The work at LANL was performed under the auspices of the U. S. DOE. ZF acknowledges support through NSF grants DMR-9870034 and DMR-9971348.

**Figure Captions**

Fig. 1. Phase diagram for $CeRhIn_5$ constructed from $C$ and $r$ [4] data (see text).

Fig. 2. (a) The specific heat, for representative values of $P$, as $C/T$ vs $T$. The $P = 0$ data are from Ref [4]. The insets show $C_e$ in the low-$T$ limit: (b) for $P > P_c$, $C_e = gT + B_2T^2$; (c) for $P < Pc$, $C_e = gT + B_{AFSW}T^3$.

Fig. 3. $C_e$ for $P \neq 0$, but with the 16 kbar data omitted for clarity.

Fig. 4. $C_e$ at 21 kbar. Normal- and superconducting-state data for $C_e$ is shown, with an extrapolation of the normal-state data to 0 K that is consistent with the superconducting-state entropy at $T_c$ and the normal-state $g$, $g(H_{c2})$. The insets show $C_e(H)$ for 50 and 70 kOe.

Fig. 5. (a) Zero-field values of $g$ vs $P$. (b) $g(H)$ vs $H$ for $P$ = 21 kbar. In (a) and (b) the open square is the 21-kbar normal-state value of $g$ obtained by the extrapolation of the 0-, 5-, and 7-T values to $H_{c2}$ = 159 kOe, represented in (b).

Fig. 6. Isotherms of $S_e(P)$ vs $P$ showing two features at 12 and 15 kbar (see text).

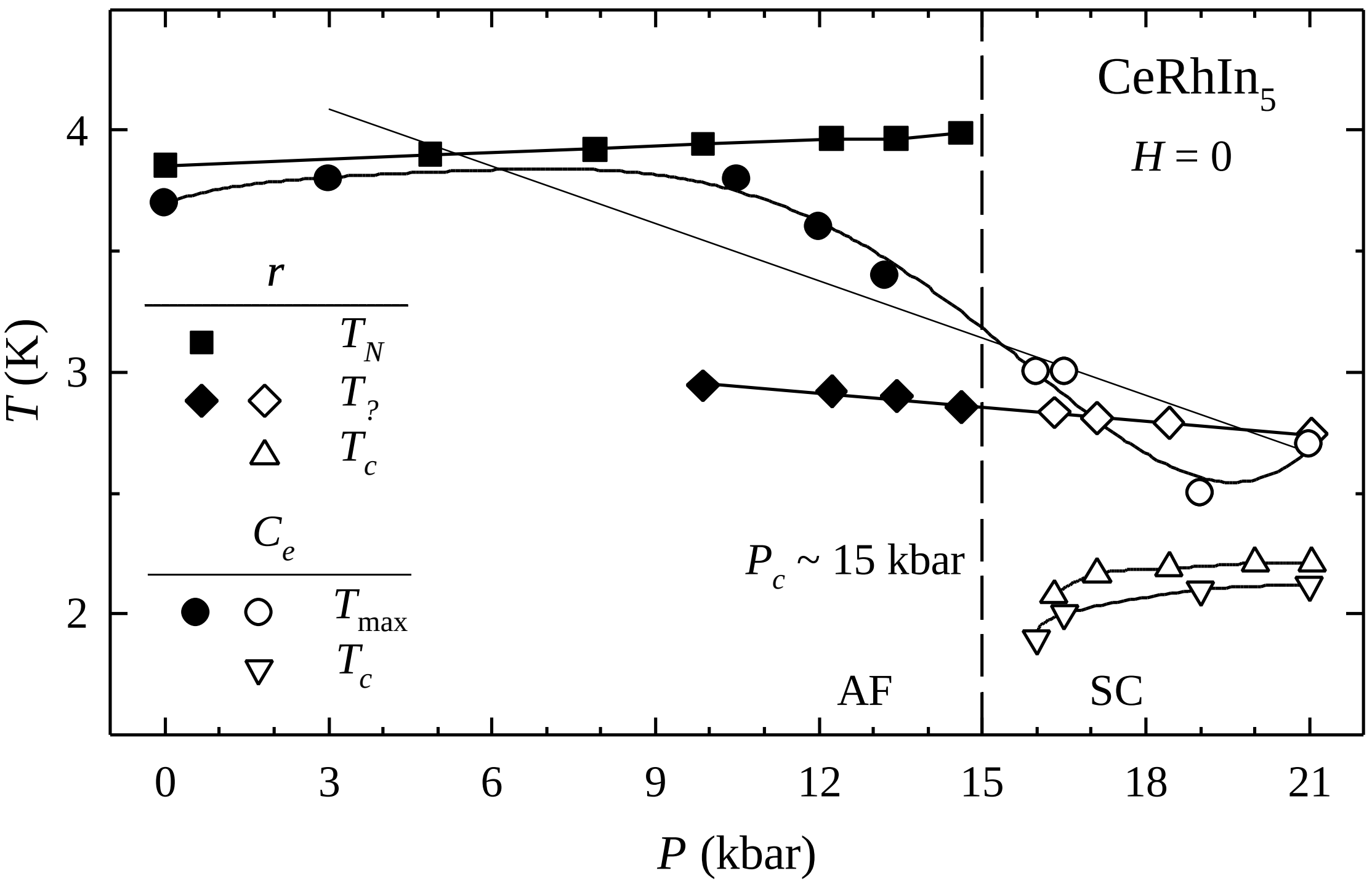

Fig. 1

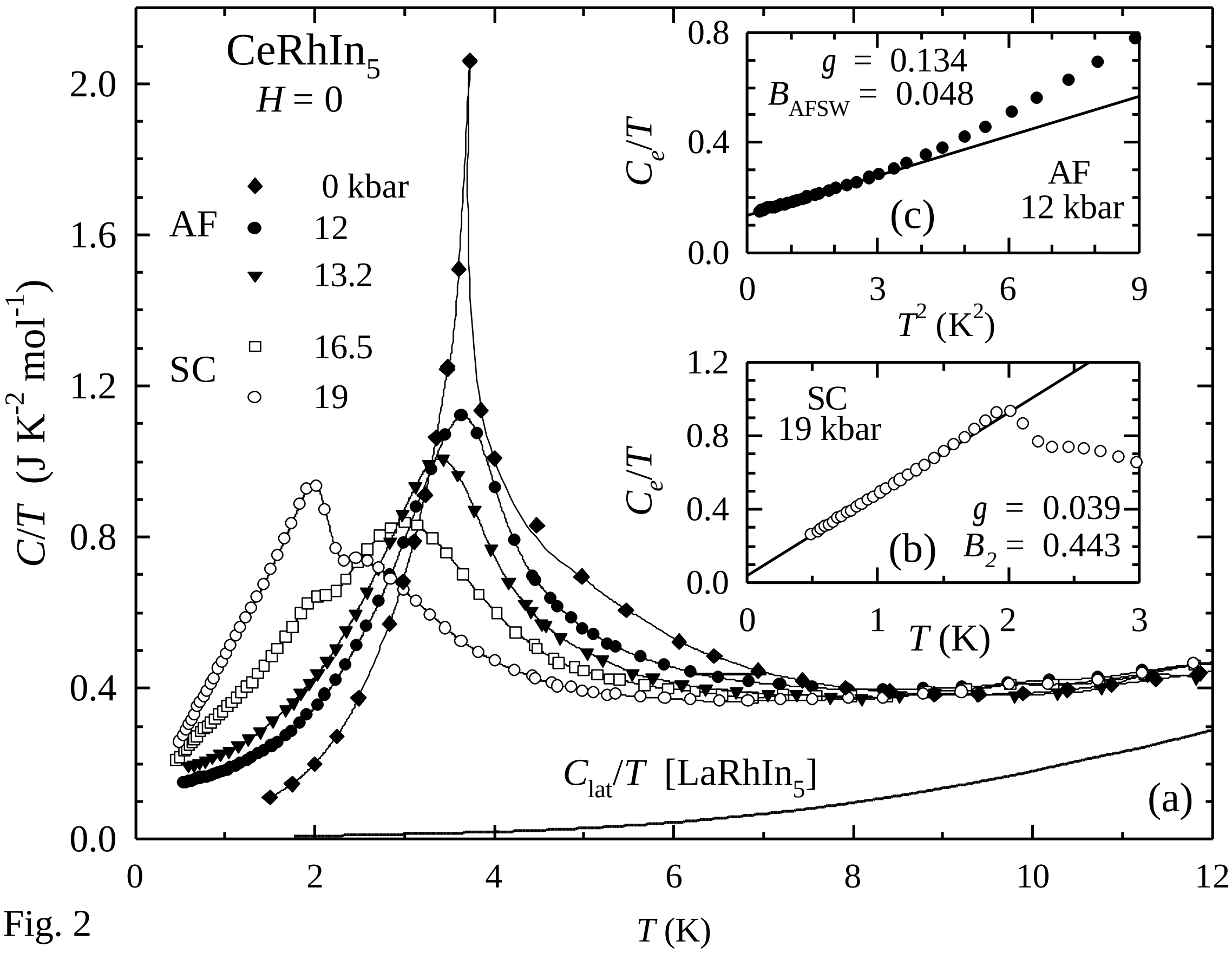

Fig. 2

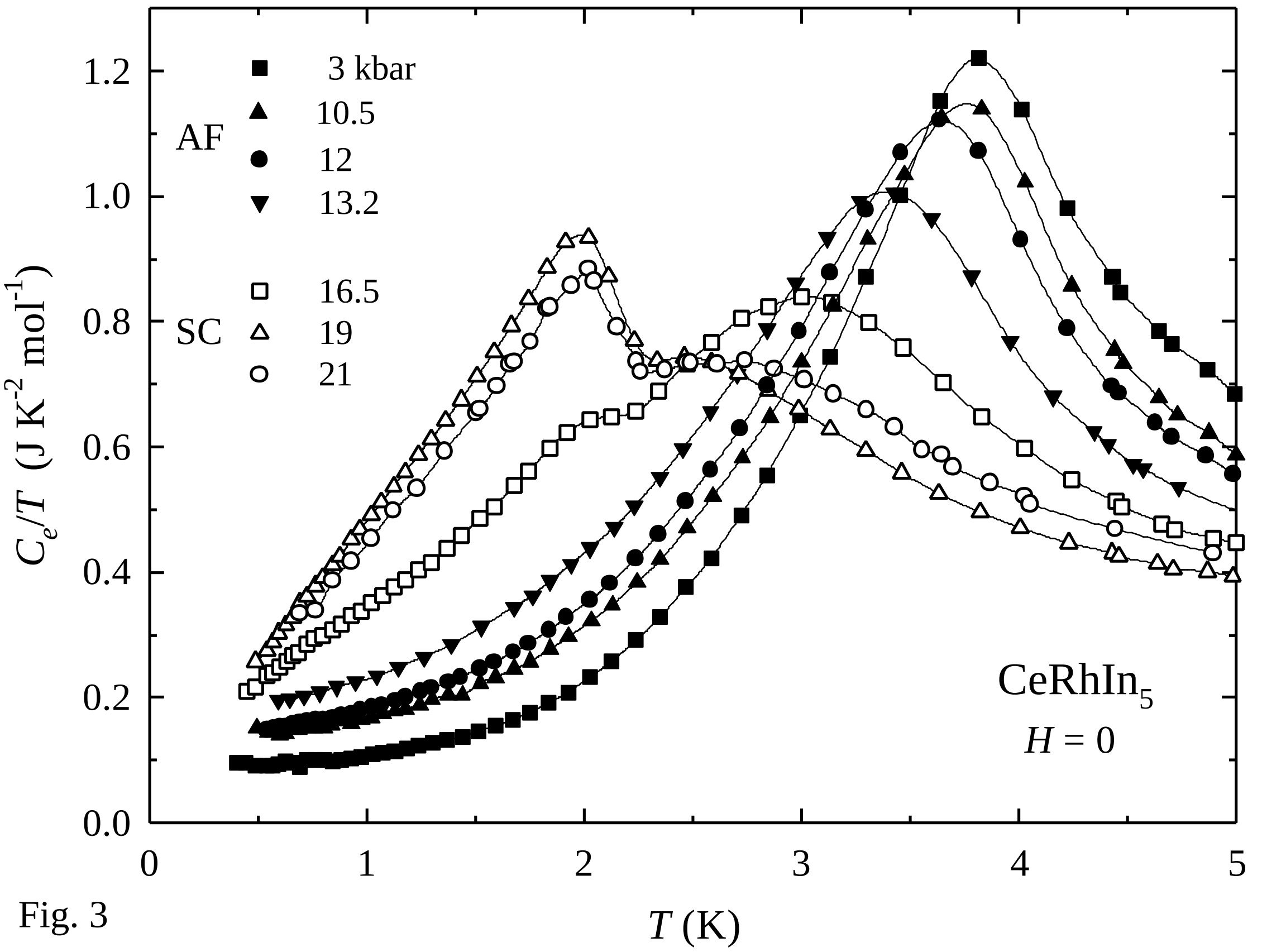

Fig. 3

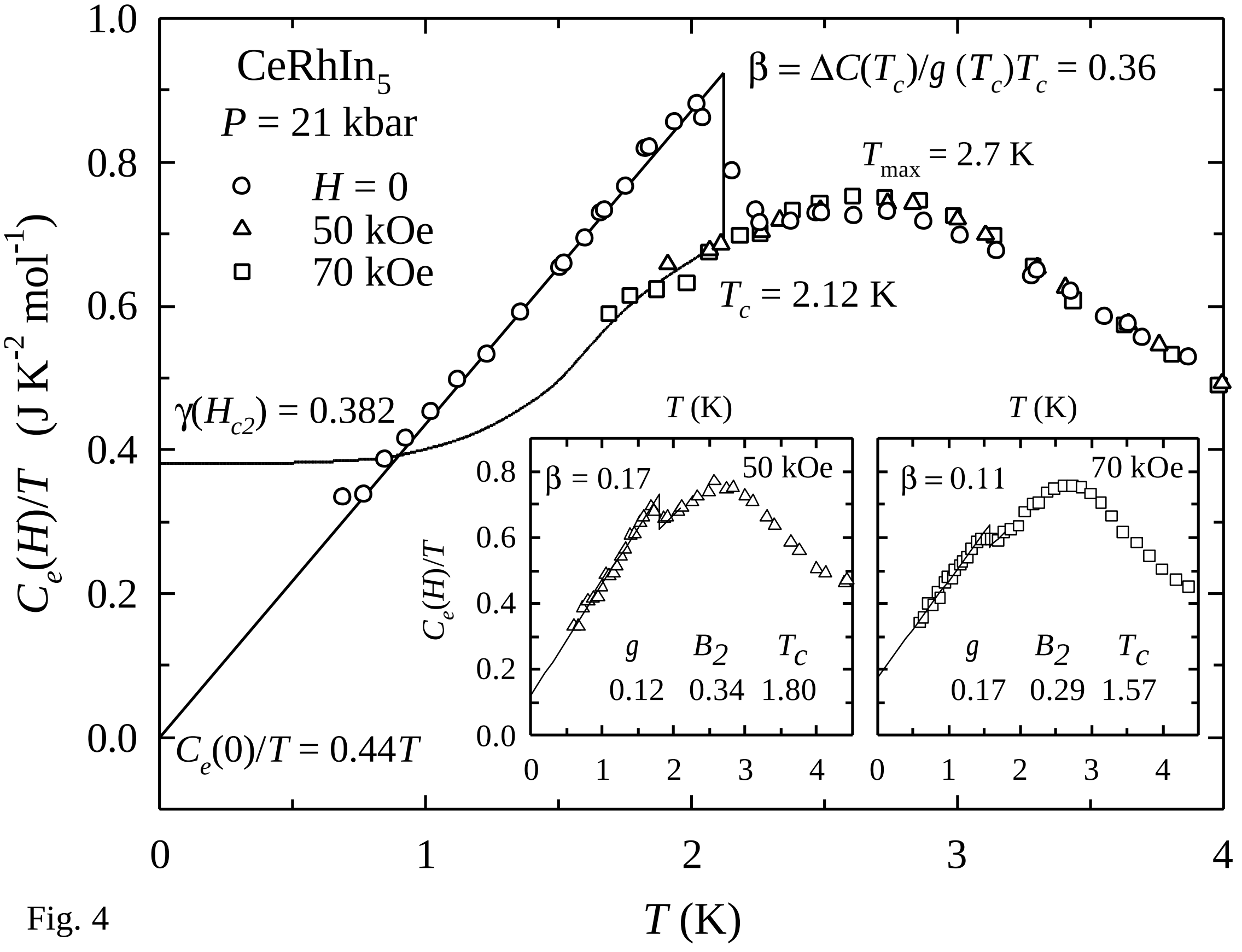
CeRhIn5
P = 21 kbar
H = 0
50 kOe
70 kOe
β = ΔC(Tc)/g (Tc)Tc = 0.36
Tmax = 2.7 K
Tc = 2.12 K
γ(Hc2) = 0.382
Ce(0)/T = 0.44T
Ce(H)/T (J K-2 mol-1)
T (K)
β = 0.17
50 kOe
g B2 Tc
0.12 0.34 1.80
β = 0.11
70 kOe
g B2 Tc
0.17 0.29 1.57
Ce(H)/T

Fig. 4

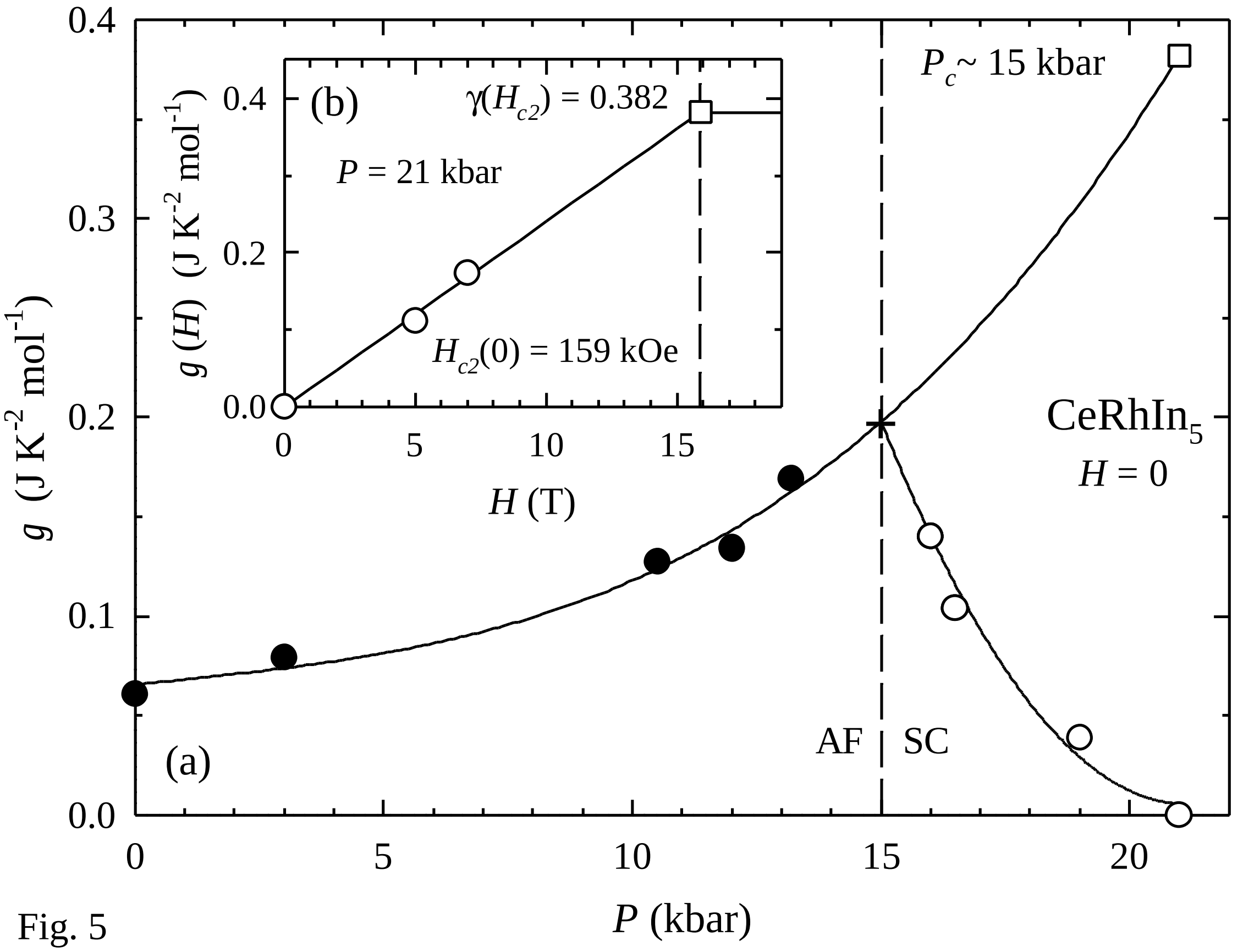

Fig. 5

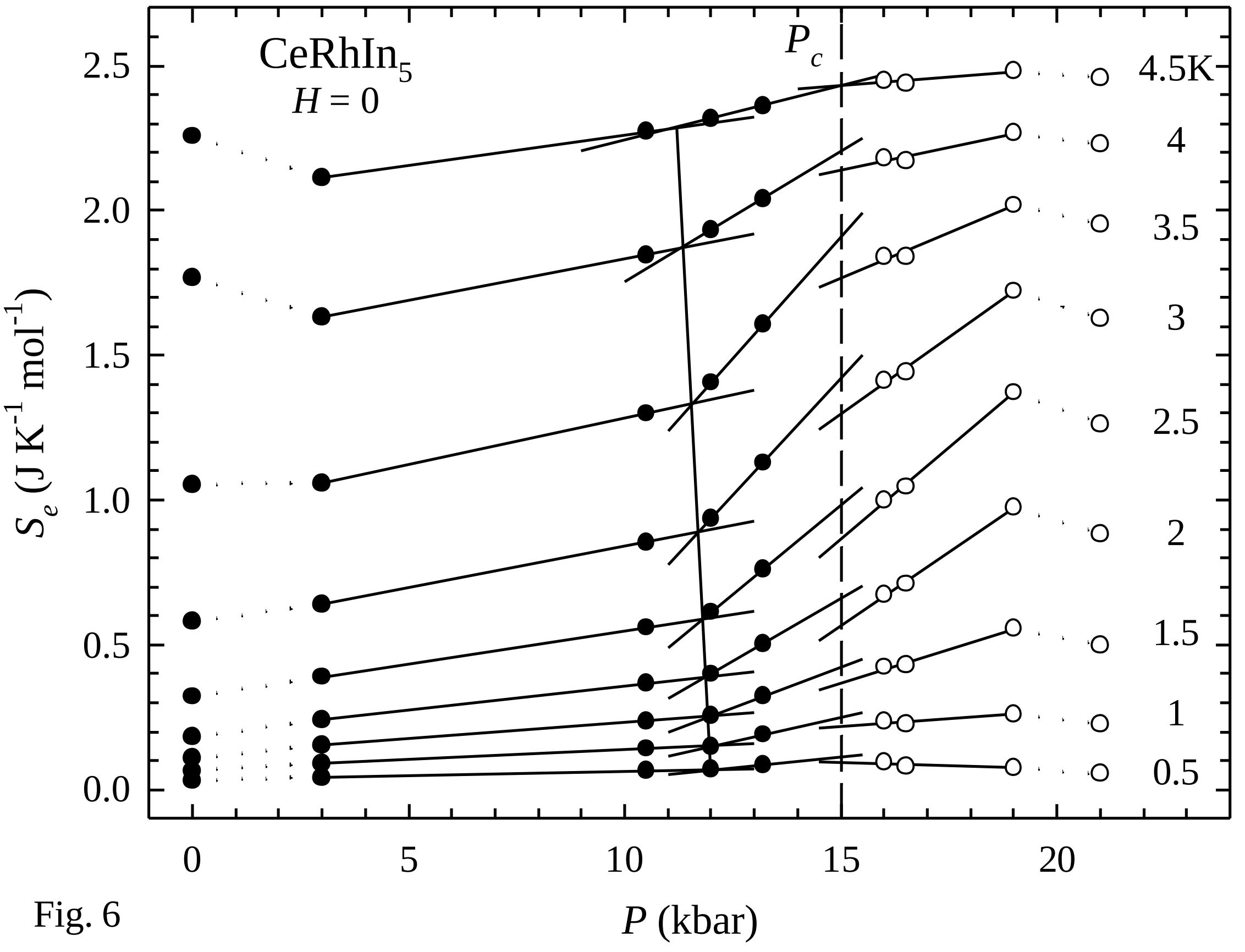

Fig. 6